\documentclass[aps,twocolumn,superscriptaddress,showpacs,showkeys,floatfix]{revtex4}
\usepackage{epsfig}
\usepackage{graphicx}
\usepackage{amsmath}
\usepackage[english]{babel}
\usepackage{multirow}
\usepackage{placeins}

\begin{document}

\title{Explosive percolation: a numerical analysis}

\author{Filippo Radicchi}
\affiliation{Complex
  Networks and Systems, ISI Foundation, Torino, Italy}
\author{Santo Fortunato}
\affiliation{Complex Networks and
  Systems, ISI Foundation, Torino, Italy}

\begin{abstract}

Percolation is one of the most studied processes in statistical physics.
A recent paper by Achlioptas {\it et al.} [{\it Science} {\bf 323}, 1453 (2009)]
has shown that the percolation transition, which is usually continuous, becomes 
discontinuous (``explosive'') if links are added to the system 
according to special cooperative rules (Achlioptas processes). In this paper we present a detailed numerical analysis
of Achlioptas processes with product rule 
on various systems, including lattices, random networks {\'a} la Erd\"os-R\'enyi and scale-free networks. 
In all cases we recover the explosive transition by Achlioptas {\it et al.}.
However, the explosive percolation transition is kind of hybrid as,
despite the discontinuity of the order parameter at the threshold, one observes 
traces of analytical behavior, like power law distributions of cluster sizes. In particular,
for scale-free networks with degree exponent $\lambda<3$, all relevant percolation variables display power law scaling,
just as in continuous second-order phase transitions.

\end{abstract}

\maketitle

\section{Introduction}

Percolation phenomena~\cite{staufferbook} represent probably the simplest examples of phase transitions that one could 
possibly imagine. On infinite lattices, the process consists in occupying sites or bonds/links with some
probability $p$. Nearest-neighboring occupied sites or links form clusters.
When $p$ exceeds a given system-dependent threshold value $p_c$, a macroscopic cluster, i.e. a cluster occupying a finite fraction of 
all available sites or links, is formed ({\it percolation cluster}). This transition is continuous, or second-order, as the order parameter
varies smoothly from zero to values greater than zero. The same type of connectedness transition not only occurs  
on regular graphs like lattices, but on any type of graphs. 
On random networks {\'a} la Erd\"os-R\'enyi (ER)~\cite{erdos}, for instance, one starts from a set of $N$ nodes and adds
links such that the probability $p$ that two nodes are joined by a link is the same for all pairs of nodes. When $p$ exceeds
the value $p_c\sim 1/N$, a percolation cluster, or {\it giant component}, emerges and the transition is again
continuous. Another well studied example is that of 
random networks with power law degree distributions of degree (number of node neighbors), 
usually called {\it scale-free} (SF) {\it networks}~\cite{Newman:2003,vitorep,barratbook}. Here the process is better defined by removing,
rather than adding, links. Links are removed until the graph is fragmented 
into microscopic clusters, i.e. there is no giant component.
Remarkably, it has been shown that, if the exponent $\lambda$ 
of the degree distribution is smaller than $3$, the giant component disappears only if one removes nearly all links of the graph, 
so that the fraction of remaining links with respect to the initial number goes to zero 
in the limit of infinite system size~\cite{cohen00,newman01,pastor00,dorogovtsev08,vazquez04}. This can be equivalently
stated by saying that the percolation threshold is zero. Nevertheless, whether the threshold is zero or non-zero,
the percolation transition is still continuous. In fact, the continuous character of the transition is 
a feature of all known percolation processes. However, this is true for {\it random percolation}, where links
are randomly placed on the system, like in the examples above. 
Recently, Achlioptas {\it et al.} have shown that, if links are placed according to special
cooperative rules, the percolation transition may become discontinuous~\cite{achlioptas09}. 
Such rules are non-local in character, as they require information between different parts of the system.
Achlioptas {\it et al.} introduced their rules in the growth of random networks, and found an abrupt jump in the size
of the giant component at the percolation threshold, hence the name {\it explosive percolation}. 
This peculiar type of transition is due to the fact that links are placed such to considerably slow down
the formation of large clusters, so that clusters are mostly 
of about the same size~\cite{friedman09,moreira09}. In this way one reaches a point in which the insertion of a vanishingly small
fraction of links leads to the merger of most of such small clusters, generating a big macroscopic cluster.
The same effect has been observed by Ziff on 2-dimensional
lattices~\cite{ziff09}. For SF networks, the problem has been studied 
by Cho {\it et al.}~\cite{cho09} and by the authors of this paper~\cite{radicchi09}, 
with different conclusions. In Ref.~\cite{cho09} the authors conclude that the 
percolation transition is discontinuous for any value of the degree distribution exponent $\lambda$ greater than
a critical value $\lambda_c \sim 2.3$; we found that, for $\lambda_c \leq \lambda \leq 3$, the behaviour at the 
percolation threshold is consistent with that of a continuous transition, while for $\lambda>3$ the expected behavior 
of a discontinuous transition is recovered. 

In this paper, we carry out an extensive numerical analysis of the phenomenon of explosive percolation.
We will describe the case of SF networks~\cite{radicchi09}, which we have studied in our previous
paper, but we will also present results on lattices and random networks. The results of random percolation  
in all graph topologies will be presented too, for comparison. 
Like in the paper by Achlioptas {\it et al.}, all graphs discussed in this paper
will be built through dynamic growth processes.
  
The paper is organized as follows. In Section~\ref{sec:models}, we describe the
growth models considered in this paper.
Section~\ref{sec:analysis} contains the results
obtained from our numerical simulations. In Section~\ref{sec:concl} we discuss the results of the analysis.
A summary is presented in Section~\ref{sec:summ}.

\section{Growth models}
\label{sec:models}

Our simulations of random percolation will be performed according to the Random Growth (RG) model, 
i.e., by iteratively adding one link to a system with $N$ nodes, where the link is 
randomly selected among all possible links. This procedure~\cite{keramiotis85,newman2000}
is equivalent to classical bond percolation. 

\begin{figure}
\begin{center}
\includegraphics[angle=-90,width=\columnwidth]{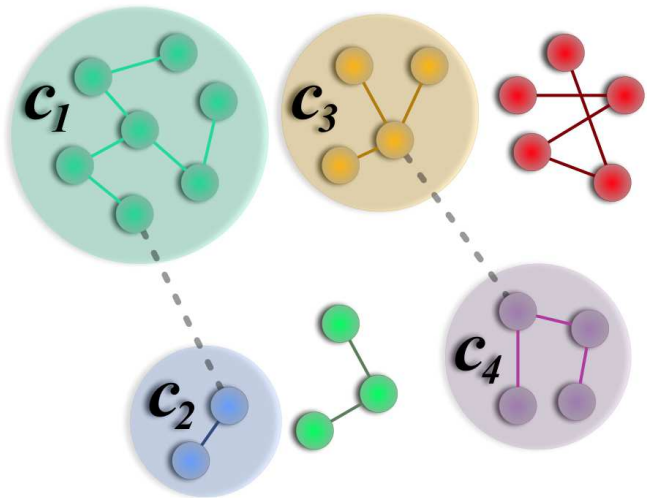}
\end{center}
\caption{(Color online) Scheme of an Achlioptas process with product rule. One of the two links represented by the dashed lines, which are
selected at random among all possible pairs of non-adjacent nodes, has to be eventually added to the system. According to 
the product rule, the winner is the link that joins the pair of clusters with the smaller product size. In this case
the winning link is that between clusters
$c_1$ and $c_2$, whose product size is $7\cdot 2=14$, which is smaller than the product of the sizes 
of clusters $c_3$ and $c_4$ ($4\cdot 4=16$).}
\label{fig1}
\end{figure}

Actually, also in Achlioptas processes links are added one by one. 
The difference is that
the link to be added is chosen among two or more randomly selected links, according to a deterministic rule. 
In this paper we focus on the Product Rule (PR), which prescribes that the link to be picked is the one minimizing
the product of the sizes of the two clusters joined by the link. The process is schematically illustrated in Fig.~\ref{fig1}.
In the paper we shall often call this specific Achlioptas process as PR model, or simply PR. 
Other options are available too. For instance one could go for the sum of the cluster sizes, instead of the product.
If, among two or more randomly selected links, the choice is random, one recovers random percolation. The systematic minimization criterion 
slows down the process of cluster growth, decelerating the percolation transition, which then may become ``explosive''.

For both RG and PR models, the growth proceeds until
one reaches the desired density of links $p$. We defined $p$ as the number of links of the graph divided by 
the total number of links present in the graph when it has been ``completed'', i.e., when the last link has been added.
All graphs considered in this paper are ``sparse'', i.e., the ratio $\langle k\rangle$ between (twice) 
the number of links and the number of nodes $N$,
which is the average degree of the nodes, does not depend on $N$. Therefore, since at time $t$ of the growth process there are exactly $t$ links in the system,
their density $p$, according to our definition, is $t/(N{\langle k\rangle})$.  

\section{Numerical analysis}
\label{sec:analysis}

Our numerical analysis aims to understand and characterize the nature of the percolation 
transition induced by RG and PR.
In order to do that, we make use of finite size scaling~\cite{landau00}, a well-known technique adopted
in numerical studies of phase transitions. For continuous phase transitions, every variable $X$ near the threshold $p_c$
is scale-independent, due to the infinite correlation length of the system at $p_c$, so it has a power law form,
\begin{equation}
X \sim |p-p_c|^{\omega},
\label{eqscal1}
\end{equation}
where $\omega$ is a critical exponent. On a finite system of size $N$, the variable $X$ has the following scaling form 
near the threshold
\begin{equation}
X = N^{-\omega/\nu}F\left[ \left(p-p_c\right)\, N^{1/\nu} \right]
\label{eqscal2}
\end{equation}
where $\nu$ is another critical exponent and $F$ a universal function.
For $p=p_c$, the variable displays the simple scaling $X \sim N^{-\omega/\nu}$, which can be used to deduce the exponents' ratio
$\omega/\nu$, from the examination of several systems with different sizes. Also, if $p_c$, $\nu$ and $\omega$ are known,
by plotting the expression $XN^{\omega/\nu}$ as a function of $\left(p-p_c\right)\, N^{1/\nu}$ one yields the universal 
function $F$, which does not depend on $N$, so curves corresponding to different system sizes collapse.

In this work we examined the two main variables measured in percolation, i.e. the {\it percolation strength} $P$
and the {\it average cluster size} $S$. The percolation strength $P$ is the order parameter of the transition, and measures
the relative size of the percolating cluster(s) with respect to the total system size $N$. On generic graphs there is no
operative criterion to define a percolating cluster (as opposed to lattices), so one usually takes $P$ as the relative size
of the largest connected cluster. The critical exponent of the percolation strength is indicated with $\beta$ and the scaling ansatz of 
$P$ is 
\begin{equation}
P = N^{-\beta/\nu}F^{(1)}\left[ \left(p-p_c\right)\, N^{1/\nu} \right].
\label{eqP}
\end{equation}
The average cluster size $S$ is defined as 
\begin{equation}
S = \frac{\sum_{s} n_s s^2}{\sum_{s}n_s s} \;,
\end{equation}
where $n_s$ stands for the number of clusters of size $s$ per node.
The sums run over all possible values of
$s$ except for the one of the largest cluster.
The critical exponent of the average cluster size is indicated with $\gamma$ and  
the scaling ansatz of 
$S$ is 
\begin{equation}
S = N^{\gamma/\nu}F^{(2)}\left[ \left(p-p_c\right)\, N^{1/\nu} \right].
\label{eqS}
\end{equation}
The universal functions $F^{(1)}$ and $F^{(2)}$ of Eqs.~(\ref{eqP}) and~(\ref{eqS}) 
are different from each other, although they are related.
We remark that in Ref.~\cite{radicchi09} we have 
used the susceptibility $\chi$ of the order parameter, which measures the size of its fluctuations, rather than the average cluster size 
$S$. Therefore, the values of $\gamma$ that we present here are different from those of Ref.~\cite{radicchi09}. 

In random percolation, the probability distribution $P(s)$ of sizes of the ``finite'' clusters, i.e. of all clusters except the largest, 
decreases as the power law $P(s)\sim s^{-\tau}$ with the system size $s$ at the percolation threshold.
In our simulations we have also measured the critical exponent $\tau$
(usually called Fisher exponent). We remark that, for a given system, $P(s)$ is proportional to $n_s$.
Their relation is $P(s)=Nn_s/n_c$, where $n_c$ is the total number of ``finite'' clusters.
This is why in the paper we shall use the symbol $n_s$ to indicate $P(s)$ as well. In the plots, however, $n_s$ is normalized as $P(s)$, 
for consistency. 

In lattice percolation, as well as in spin models, the exponents $\beta_L$, $\gamma_L$ and $\nu_L$ (where $L$ stays
for lattice) are linked by the 
so-called {\it hyperscaling relation}
\begin{equation}
\frac{\gamma_L}{\nu_L}+\frac{2\beta_L}{\nu_L}=d,
\label{hyper}
\end{equation}
where $d$ is the dimension of the lattice. In the general case of graphs, we do not have a space dimension, so the 
scaling is done in terms of the ``volume'' $N$, as we have done in Eqs.~(\ref{eqscal2}),~(\ref{eqP}) and~(\ref{eqS}). 
In lattices $N=L^d$ and the hyperscaling relation for the exponents
expressing the scaling of the variables in terms of the volume $N$ reads
\begin{equation}
\frac{\gamma}{\nu}+\frac{2\beta}{\nu}=1.
\label{hyper1}
\end{equation}

Eq.~(\ref{hyper1}) is actually very general, and holds for random percolation on any system below the 
upper critical dimension~\cite{cohen02}.

The identification of the percolation threshold $p_c$ is performed in two independent ways.
One way consists in using the scaling of the pseudo-critical points $p_c(N)$
\begin{equation}
p_c = p_c(N) + b N^{-1/\nu} \,,
\label{eq:chi2}
\end{equation}
where $b$ is a constant which has to be determined from the fit 
together with the other parameters $\nu$ and $p_c$. The pseudocritical points can be defined in several ways. 
We took the positions of the peaks of $S$ at different system sizes $N$.

The second method is based on Eq.~(\ref{eqP}). By plotting the percolation 
strength $P$ as a function of the system size $N$ for a given value of $p$, the correct
value of the percolation threshold can be determined by finding the value of $p$ which yields the best power law fit.

In~\cite{achlioptas09} a new method for the determination of the nature of the transition has been proposed. 
The method consists in studying the behaviour of the width of the transition window as a function of the system size. 
As a measure of the width of the transition window 
we considered the quantity $\Delta p=p_2-p_1$, where $p_2$ is the lowest value of $p$ for which $P > 0.5$ and 
$p_1$ the lowest value of $p$ for which $P > 1/\sqrt{N}$. As we will see, the width of the transition window generally 
scales as a power law with the system size and its dependence from $N$ can be therefore written as $\Delta p \sim N^{-\alpha}$. 
Achlioptas {\it et al.} argued that, for continuous transitions, $\Delta p$ should be independent of the system size ($\alpha=0$), whereas,
if there is an explosive first-order transition, $\Delta p$ should decrease with $N$ ($\alpha>0$). Actually, in 2-dimensional lattices 
Ziff has found that $\alpha>0$ even in the case of random percolation~\cite{ziff09}. This is however due to the fact that in the particular
case of the lattice $p_2$ is essentially
coincident with the actual critical threshold of the system; therefore on the lattice one should take a value $p_3$ appreciably 
larger than $p_c$ (like the point at which $P>0.7$, for instance). We stress that
the choice of $p_1$ and $p_2$ is completely arbitrary, so the robustness of the exponent $\alpha$ needs to be tested.
Therefore we also used another definition of $\Delta p$, namely $\Delta\tilde{p}=\tilde{p}_2-p_1$, 
where $\tilde{p}_2$ is the lowest value of $p$ for which 
$P > 0.2$. Also in this case, we can generally write  $\Delta \tilde{p} \sim N^{-\tilde{\alpha}}$. 
The robustness of the scaling of $\Delta p$ would be indicated by the equality of the exponents $\alpha$  and $\tilde{\alpha}$. 

\subsection{Lattices}
\label{sec:2d}

\begin{figure}[htb]
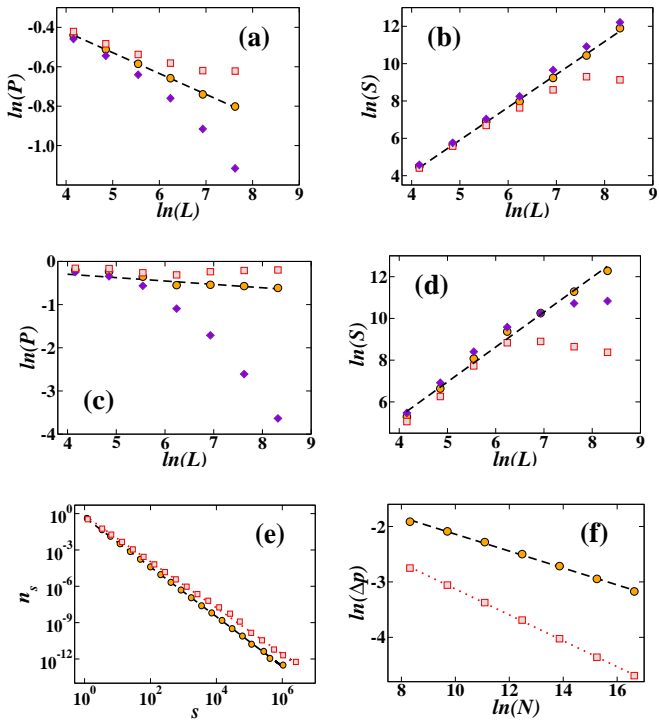

\includegraphics[width=0.22\textwidth, height=0.16\textwidth]{beta_d2_cl}
\qquad 
\includegraphics[width=0.22\textwidth,  height=0.16\textwidth]{gamma_d2_cl}

\vskip .4cm

\includegraphics[width=0.22\textwidth, height=0.16\textwidth]{beta_d2_pr}
\quad 
\includegraphics[width=0.22\textwidth,  height=0.16\textwidth]{gamma_d2_pr}

\vskip .4cm

\includegraphics[width=0.22\textwidth,  height=0.16\textwidth]{tau_d2.eps}
\quad 
\includegraphics[width=0.22\textwidth,  height=0.16\textwidth]{ac_d2_.eps}

\caption{(Color online) Analysis of $2d$-lattices. (a) RG model: the percolation strength $P$ is plotted as a function of the lattice 
side $L$ for three different values of the occupation probability: $p=0.499$ (violet diamonds), $p=0.5$ (orange circles) 
and $p=0.501$ (grey squares). The dashed line stands for the best fit obtained at the critical point $p=p_c=0.5$, which 
allows to determine $\beta/\nu=0.11(1)$. (b) RG model: the average cluster size $S$ is plotted as a function of the 
lattice side $L$ for the same values of $p$ as those used in (a). The dashed line has slope $\gamma/\nu=1.76(1)$. 
(c) PR model: the percolation strength $P$ is plotted as a function of the lattice side $L$ for three different 
values of the occupation probability: $p=0.5256$ (violet diamonds), $p=0.5266$ (orange circles) and $p=0.5276$ 
(grey squares). The dashed line stands for the best fit obtained at the critical point $p=p_c=0.5266(2)$, which allows 
to determine $\beta/\nu=0.07(3)$. (d) PR model: $S$ is plotted as a function of the lattice side $L$ for the same 
values of $p$ used in (c). The dashed line has slope $\gamma/\nu=1.7(1)$. (e) Comparison between the cluster 
size distributions measured at the critical threshold for both growth models. For RG (orange circles) $\tau=2.05(1)$ 
(black dashed line), while for PR (grey squares) $\tau=1.9(1)$ (red dotted line). Simulations have been 
performed on systems with $L=4096$. (f) $\Delta p$ as a function of the system size $N$: $\alpha=0.15(1)$ 
(dashed black line) for RG (orange circles) and $\alpha= 0.24(1)$ (dotted red line) for PR (grey squares). The first value is questionable,
as the scaling should yield a plateau ($\alpha\sim 0$), like we have indirectly verified (see text). To see the correct scaling one should 
simulate much larger systems.}
\label{fig:d2}
\end{figure}

We consider first the case of $2d$-lattices (square lattices) with periodic boundary conditions.
The results obtained from our simulations with RG 
(see Figure~\ref{fig:d2}) confirm the well-known fact that the transition is continuous. We also recover the correct 
critical exponents; the scaling is done in terms
of the linear dimension $L$ of the lattice, as it is customary in this case. 
PR on $2d$-lattices has been only recently studied by Ziff~\cite{ziff09}, 
who has shown that the transition is explosive, like that observed by Achlioptas. 
In Figs.~\ref{fig:d2}c,~\ref{fig:d2}d,~\ref{fig:d2}e and \ref{fig:d2}f we report the results 
obtained by applying PR on $2d$-lattices. We find a trivial 
scaling for the order parameter $P$, with exponents' ratio $\beta/\nu$ basically equal to zero [$0.07(3)$] (Fig.~\ref{fig:d2}c). This is consistent
with what one expects to find for a discontinuous transition. 
On the other hand, we find a clean non-trivial power law scaling at $p_c$ for the average cluster size $S$, with exponent
$\gamma/\nu=1.7(1)$ (Fig.~\ref{fig:d2}d). This 
had been observed by Cho {\it et al.} in SF networks~\cite{cho09}. An explanation of this is provided by Fig.~\ref{fig:d2}e, which shows
the distribution of sizes $n_s$ for all clusters except the largest one. The distribution is a clear power law 
[exponent $1.9(1)$], which is unexpected for 
a classic discontinuous transition, as it usually occurs for continuous transitions. 
Therefore, all variables derived from $n_s$, like the average cluster size $S$, display power law scaling.
This striking feature, as we will see below, is common to all ``explosive'' transitions we have investigated here.
\FloatBarrier
Finally, in Fig.~\ref{fig:d2}f we show the results of the Achlioptas test for both RG and PR. For RG, we find $\alpha=0.15(1)$. 
As we remarked above, $\alpha$ is non-zero despite the continuous percolation transition, which seems to go against the argument
by Achlioptas {\it et al.}. However, this happens because $p_2$ is very close to the critical point $p_c$. The correct behavior
can be seen if one considers a window clearly including $p_c$, which could be done by taking a larger value for the upper limit of the window,
like, e.g. the smallest value $p_3$ at which the relative size of the giant component exceeds $0.7$. Actually the scaling of 
$p_3-p_1$ (not shown) still shows sublinear behavior, but we believe that this is due to the fact that $p_1$ grows too rapidly for the 
systems we were able to simulate. In fact, $p_3-p_2$ is approximately constant for the lattice sizes we have taken, 
so $p_3-p_1>p_3-p_2$ cannot go to zero in the limit of infinite lattice size.
For PR, we obtain $\alpha=0.24(1)$. This result is quite different from the value found by Ziff ($0.34$).
However, in his simulations, Ziff has considered only links between clusters, whereas we have 
considered all possible links, including those within clusters. Simulations of the process \'a la Ziff has confirmed that 
this is indeed the reason of the discrepancy with our result. We have performed the same analysis for the window 
$\Delta\tilde{p}$ defined in Section~\ref{sec:analysis}: the exponents $\tilde{\alpha}$ for RG and PR are consistent
with the corresponding values of $\alpha$ (see Table~\ref{table}).

\begin{figure}
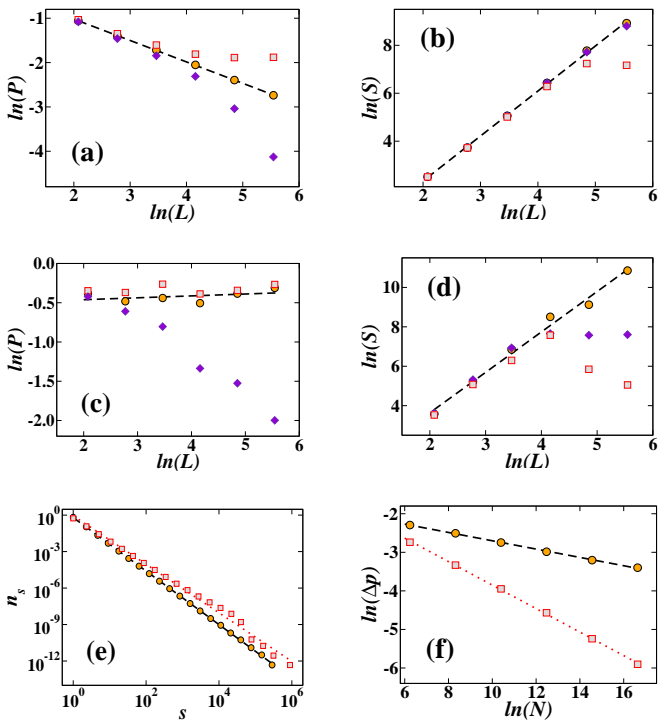


\includegraphics[width=0.22\textwidth,  height=0.16\textwidth]{beta_d3_cl}
\qquad
\includegraphics[width=0.22\textwidth,  height=0.16\textwidth]{gamma_d3_cl}

\vskip .4cm

\includegraphics[width=0.22\textwidth,  height=0.16\textwidth]{beta_d3_pr}
\qquad
\includegraphics[width=0.22\textwidth,  height=0.16\textwidth]{gamma_d3_pr}

\vskip .4cm

\includegraphics[width=0.22\textwidth,  height=0.16\textwidth]{tau_d3.eps}
\qquad
\includegraphics[width=0.22\textwidth,  height=0.16\textwidth]{ac_d3_.eps}

\caption{(Color online) Analysis of $3d$-lattices. (a) RG model: the percolation strength $P$ is plotted as a function 
of the lattice side $L$ for three different values of the occupation probability: $p=0.2478$ (violet diamonds), 
$p=0.2488$ (orange circles) and $p=0.2498$ (grey squares). The dashed line stands for the best fit obtained 
at the critical point $p=p_c=0.2488(3)$, which allows to determine $\beta/\nu=0.48(1)$. (b)  RG model:  the average 
cluster size $S$ is plotted as a function of the lattice side $L$ for the same values of $p$ used in (a). 
The dashed line has slope $\gamma/\nu=2.0(1)$. (c) PR model: the percolation strength $P$ is plotted as a 
function of the lattice side $L$ for three different values of the occupation probability: $p=0.3866$ 
(violet diamonds), $p=0.3876$ (orange circles) and $p=0.3886$ (grey squares). The dashed line stands for 
the best fit obtained at the critical point $p=p_c=0.3876(2)$, which allows to determine $\beta/\nu=0.02(2)$. 
(d) PR model: the average cluster size $S$ is plotted as a function of the lattice side $L$ for the same 
values of $p$ used in (c). The dashed line has slope $\gamma/\nu=2.1(1)$. (e) Cluster size distributions 
$n_s$ for $3d$-lattices at the critical point. For both percolation models $n_s \sim s^{-\tau}$  as $s$ increases. 
For RG (orange circles) $\tau=2.20(1)$ (black dashed line), 
while for PR (grey squares) $\tau=1.99(4)$ (red dotted line). Simulations 
have been performed on systems with $L=256$. (f)  We plot the quantity $\Delta p$ as a function of the 
system size $N$. For both models $\Delta p$ decreases as a power law, $\Delta p \sim N^{-\alpha}$, 
as $N$ increases. In particular we have: $\alpha=0.10(1)$ (dashed black line) for RG (orange circles) 
and $\alpha=0.30(1)$ (dotted red line) for PR (grey squares). The first value is questionable,
as the scaling should yield a plateau ($\alpha\sim 0$), like we have indirectly verified (see text). To see the correct scaling one should 
simulate much larger systems.}
\label{fig3d}
\end{figure}

In 3d-lattices, the general picture is consistent with that in two dimensions (Fig.~\ref{fig3d}). 
Classic percolation results, threshold and exponents, are recovered (Fig.~\ref{fig3d}a,~\ref{fig3d}b,~\ref{fig3d}e). 
The scaling at $p_c$ of the order parameter $P$ for the PR process is again trivial,
with exponent $\beta/\nu=0.02(2)$, essentially zero (Fig.~\ref{fig3d}c). 
The $S$ scales with an exponent $\gamma/\nu=2.1(1)$ (Fig.~\ref{fig3d}d), again due to the power law shape 
of the distribution of cluster sizes (Fig.~\ref{fig3d}e). We remark that the exponent $\tau=1.99(4)$ is compatible with
that we found in two dimensions [$1.9(1)$].
The test of Achlioptas {\it et al.} (Fig.~\ref{fig3d}f) yields again
a non-zero value of the exponent $\alpha$ for RG ($\alpha=0.10(1)$) (probably because our lattices are not yet large enough to see
the actual behavior, as in $2d$), and a larger value for PR ($\alpha=0.30(1)$).
Like in two dimensions, also in $3d$ the exponents $\tilde{\alpha}$ for RG and PR are consistent
with the corresponding values of $\alpha$ (see Table~\ref{table}).

\subsection{Erd\"os-R\'enyi random networks}
\label{sec:er}

\begin{figure}
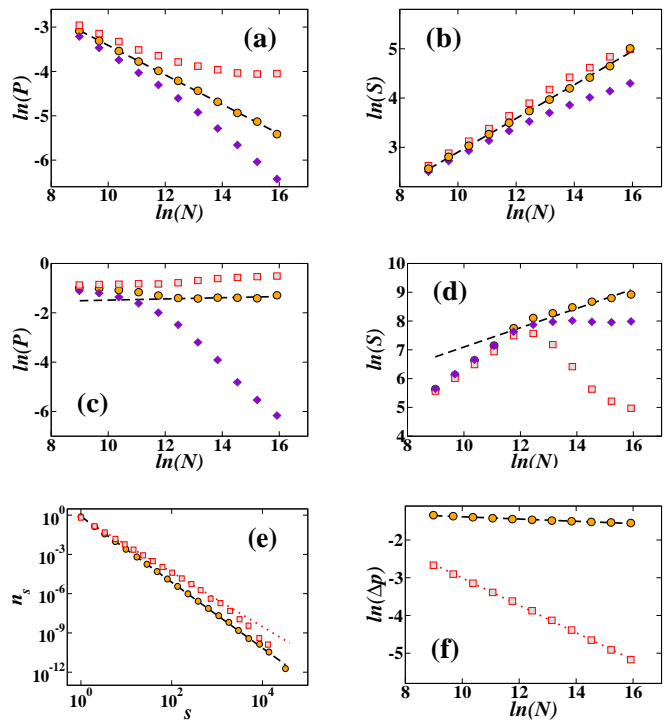

\includegraphics[width=0.22\textwidth,  height=0.16\textwidth]{beta_er_cl}
\qquad
\includegraphics[width=0.22\textwidth,  height=0.16\textwidth]{gamma_er_cl}

\vskip .4cm

\includegraphics[width=0.22\textwidth,  height=0.16\textwidth]{beta_er_pr.eps}
\qquad
\includegraphics[width=0.22\textwidth,  height=0.16\textwidth]{gamma_er_pr.eps}

\vskip .4cm

\includegraphics[width=0.22\textwidth,  height=0.16\textwidth]{tau_er.eps}
\qquad
\includegraphics[width=0.22\textwidth,  height=0.16\textwidth]{ac_er_.eps}

\caption{(Color online) Analysis of ER random networks. (a) RG model: the percolation strength $P$ is plotted as a function of 
the system size $N$ for three different values of the occupation probability: $p=0.495$ (violet diamonds), 
$p=0.5$ (orange circles) and $p=0.505$ (grey squares). The dashed line stands for the best fit obtained at the 
critical point $p=p_c=0.5$, which allows to determine $\beta/\nu=0.33(1)$. (b) RG model: the average cluster size 
$S$ is plotted as a function of the network size $N$ for the same values of $p$ used in (a). The dashed line 
has slope $\gamma/\nu=0.34(1)$. (c) PR model: the percolation strength $P$ is plotted as a function of the 
network size $N$ for three different values of the occupation probability: $p=0.8872$ (violet diamonds), 
$p=0.8882$ (orange circles) and $p=0.8892$ (grey squares). The dashed line stands for the best fit obtained 
at the critical point $p=p_c=0.8882(2)$, which allows to determine $\beta/\nu=0.02(1)$. (d) PR model: 
the average cluster size $S$ is plotted as a function of the network size $N$ for the same values of $p$ used in (c). 
The dashed line has slope $\gamma/\nu=0.48(4)$. (e) Cluster size distributions $n_s$ for ER random networks at critical point. 
For both percolation models $n_s \sim s^{-\tau}$  as $s$ increases. For RG (orange circles) 
$\tau=2.51(2)$ (black dashed line), while for PR (grey squares)
$\tau=2.08(5)$ (red dotted line). Simulations have been performed on systems with $N=8192$. (f) We plot 
the quantity $\Delta p$ as a function of the system size $N$. For both models $\Delta p$ 
decreases as a power law, $\Delta p \sim N^{-\alpha}$, as $L$ increases. In particular we have: $\alpha=0.03(1)$ 
(dashed black line) for RG (orange circles) and $\alpha=0.36(1)$ (dotted red line)  
for PR (grey squares).}
\label{figER}
\end{figure}

Percolation studies on random networks {\'a} la Erd\"os-Ren\'yi (ER) have a long tradition, as we wrote in the Introduction. 
Fig.~\ref{figER} summarizes the results of our analysis. The well-known results of random percolation, threshold and exponents,
are recovered, as illustrated in Figs.~\ref{figER}a,~\ref{figER}b and \ref{figER}e. In particular, we notice that the 
hyperscaling relation of Eq.~\ref{hyper1} is satisfied for the exponents' ratios $\beta/\nu$ and $\gamma/\nu$. For PR,
instead, we see again a flat profile of the order parameter $P$ with $N$
($\beta/\nu=0.02(1)$), which hints to a discontinuous transition, together with a power law scaling of the average
cluster size $S$, with exponent $\gamma/\nu=0.48(4)$. The exponent $\tau=2.08(5)$ (Fig.~\ref{figER}e) is still compatible with the values found for 
PR on both $2d$ and $3d$ lattices (see Section~\ref{sec:2d} and Table~\ref{table}). The Achlioptas test of Fig.~\ref{figER}f
yields $\alpha=0.03(1)$ for RG, compatible with a window $\Delta p$ that is independent of $N$, while for PR 
$\alpha=0.36(1)$, in agreement with the calculations of Achlioptas {\it et al.}~\cite{achlioptas09}. Again, the same test performed 
with the window $\Delta\tilde{p}$ yields essentially the same values of the exponent for both RG and PR (Table~\ref{table}), 
so the results of the test appear to be quite robust.

\subsection{Scale-free networks}
\label{sec:sf}

\begin{figure}
\begin{center}
\includegraphics[width=\columnwidth]{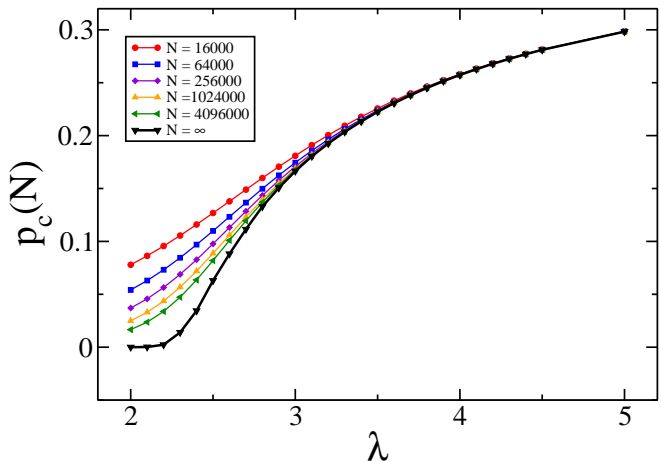}
\end{center}
\caption{(Color online) Achlioptas process with PR on random SF networks. The plot shows the percolation
threshold $p_c(N)$ as a function of the degree exponent $\lambda$ for various network sizes $N$. 
The black line represents the infinite size limit extrapolation of the critical threshold, performed by applying 
Eq.~(\ref{eq:chi2}). The percolation threshold becomes non-zero for $\lambda>\lambda_c\sim 2.3$. Reprinted figure 
with permission from Ref.~\cite{radicchi09}.}
\label{figthres}
\end{figure}
\begin{figure}
\includegraphics[width=0.22\textwidth,  height=0.16\textwidth]{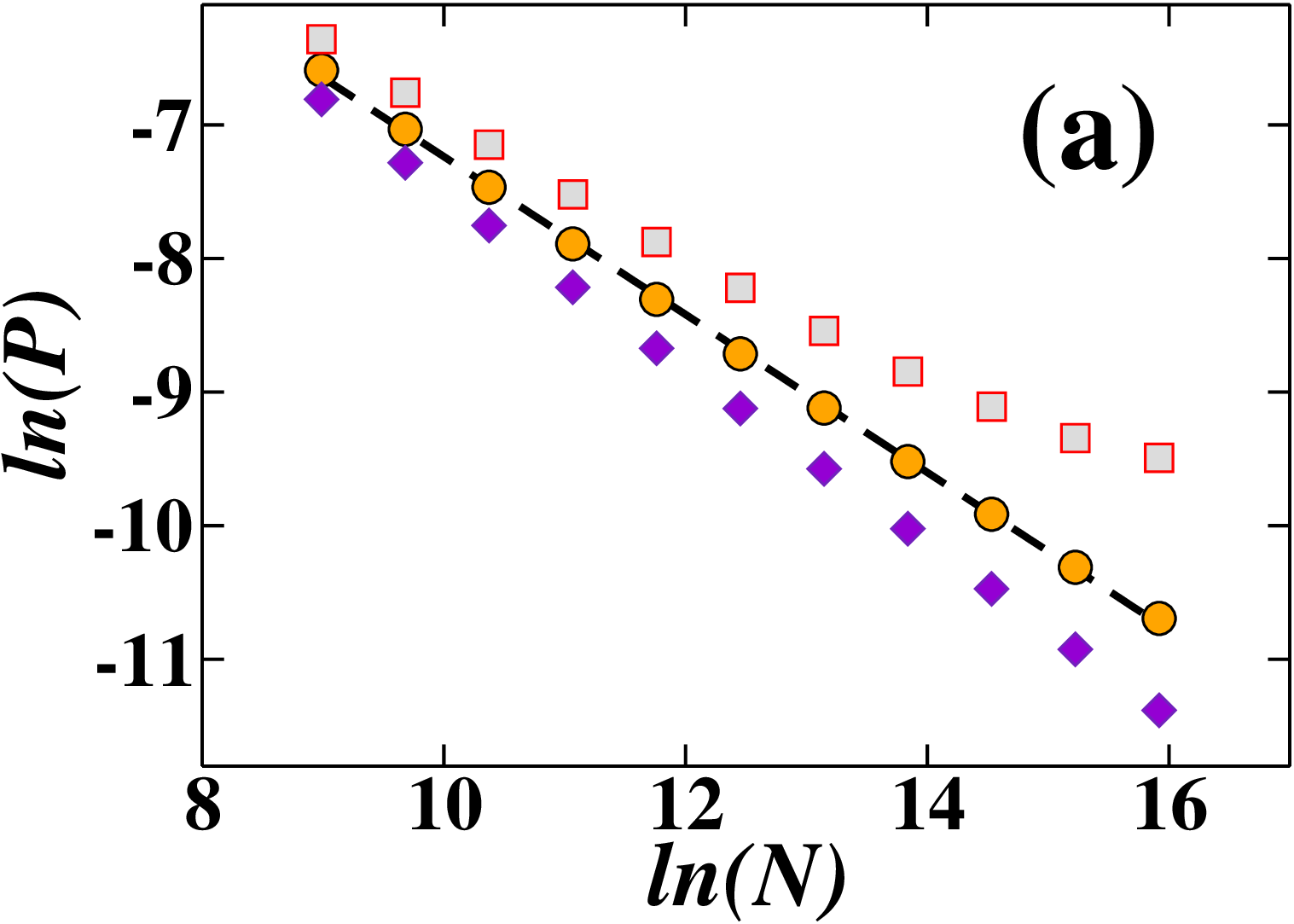}
\qquad
\includegraphics[width=0.22\textwidth,  height=0.16\textwidth]{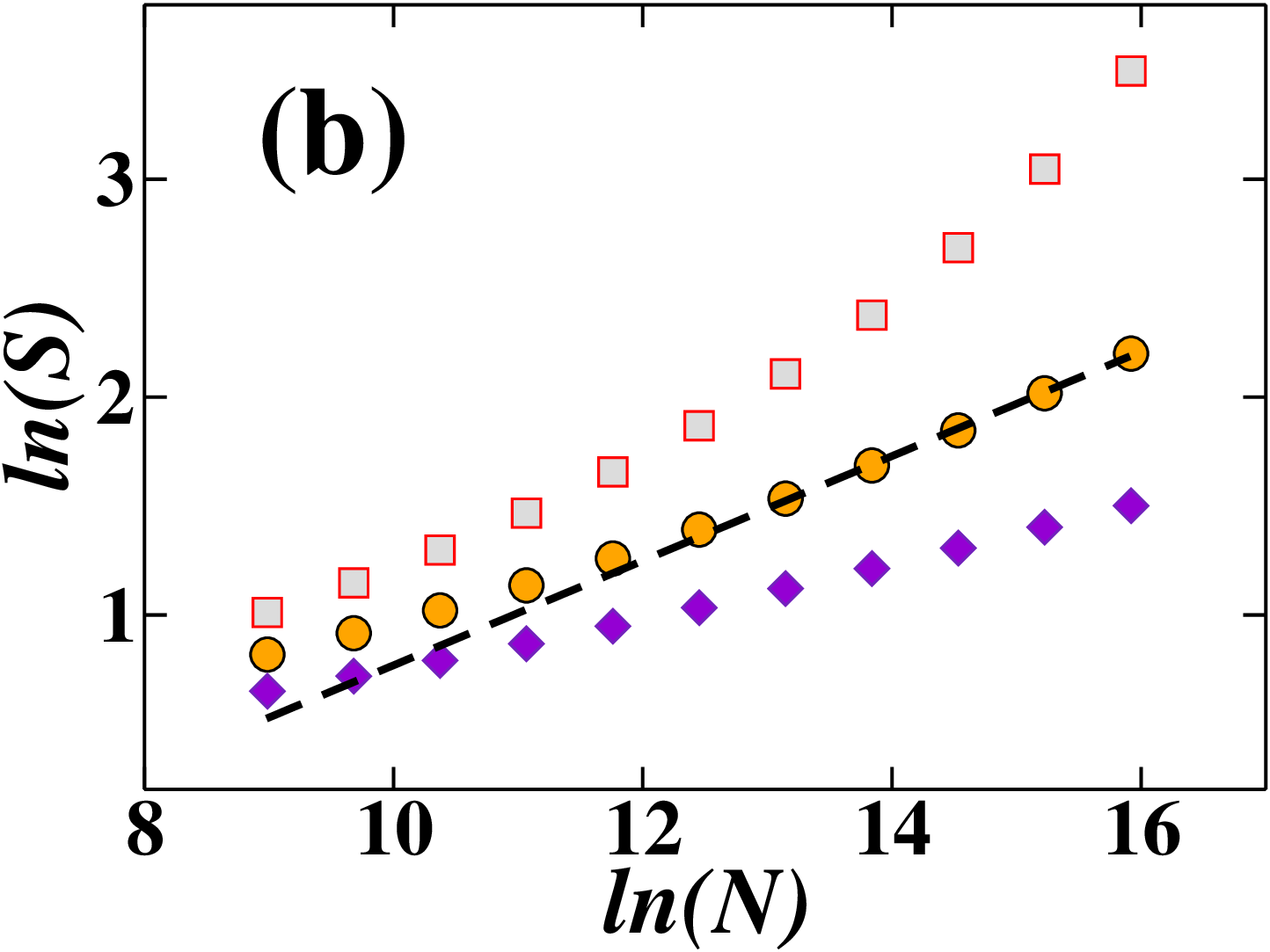}
\caption{(Color online) Percolation transition induced by an Achlioptas process with PR on SF networks. The degree exponent $\lambda=2.5$. 
(a) The percolation strength $P$ is plotted as a function of the system size $N$ for three 
different values of the occupation probability: $p=0.0529$ (violet diamonds), $p=0.0629$ 
(orange circles) and $p=0.0729$ (grey squares). The dashed line stands for the best fit 
obtained at the critical point $p=p_c=0.0629(1)$, which allows to determine $\beta/\nu=0.59(1)$. 
(b) The average cluster size $S$ is plotted as a function of the network size $N$ for the same 
values of $p$ used in (a). The dashed line has slope $\gamma/\nu=0.24(1)$.}
\label{figSF2.5}
\end{figure}
\begin{figure}
\includegraphics[width=0.22\textwidth,  height=0.16\textwidth]{beta_sf_g2.8_pr.eps}
\quad
\includegraphics[width=0.22\textwidth,  height=0.16\textwidth]{gamma_sf_g2.8_pr.eps}
\caption{(Color online) Percolation transition induced by an Achlioptas process with PR on SF networks.
The degree exponent $\lambda=2.8$. (a) The 
percolation strength $P$ is plotted as a function of the system size $N$ for three 
different values of the occupation probability: $p=0.1229$ (violet diamonds), $p=0.1329$ 
(orange circles) and $p=0.1349$ (grey squares). The dashed line stands for the 
best fit obtained at the critical point $p=p_c=0.1329(1)$, which allows to determine 
$\beta/\nu=0.50(1)$. (b)  The average cluster size $S$ is plotted as a function of 
the network size $N$ for the same values of $p$ used in (a). The dashed line has slope $\gamma/\nu=0.42(1)$.}
\label{figSF2.8}
\end{figure}
\begin{figure}
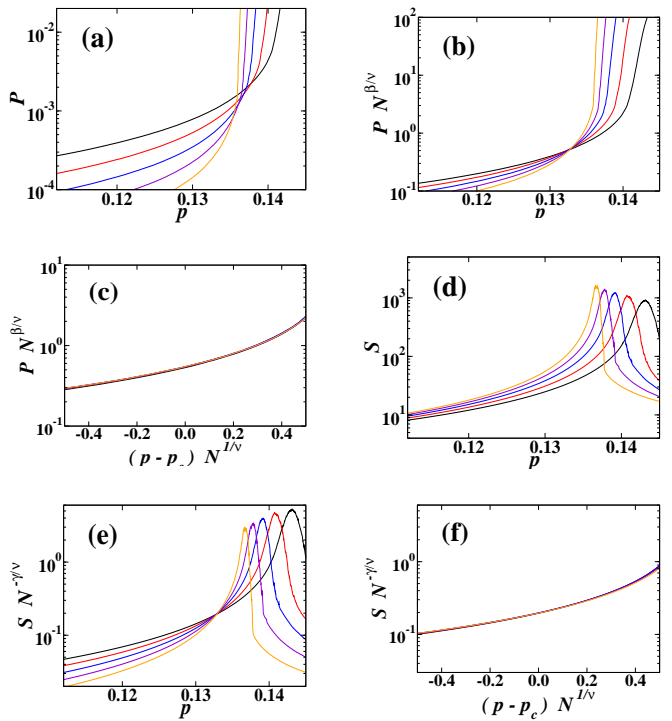

\includegraphics[width=0.22\textwidth, height=0.16\textwidth]{scaling1_sf_g2.8_pr.eps}
\qquad 
\includegraphics[width=0.22\textwidth,  height=0.16\textwidth]{scaling2_sf_g2.8_pr.eps}

\vskip 0.4cm

\includegraphics[width=0.22\textwidth, height=0.16\textwidth]{scaling3_sf_g2.8_pr.eps}
\qquad 
\includegraphics[width=0.22\textwidth,  height=0.16\textwidth]{g_scaling1_sf_g2.8_pr.eps}

\vskip 0.4cm

\includegraphics[width=0.22\textwidth,  height=0.16\textwidth]{g_scaling2_sf_g2.8_pr.eps}
\qquad 
\includegraphics[width=0.22\textwidth,  height=0.16\textwidth]{g_scaling3_sf_g2.8_pr.eps}

\caption{(Color online) Rescaling of the percolation variables $P$ and $S$ near the percolation transition 
induced by an Achlioptas process with 
PR on SF networks with degree exponent $\lambda=2.8$. System size $N$ goes from $256000$ to $4096000$ via successive doublings.}
\label{fig:scaling_sf2.8}
\end{figure}

SF networks have been objects of intense investigations over the last few years~\cite{Newman:2003,vitorep,barratbook}.
The main reason of their success 
is that they are a proxy of many natural, social and man-made systems, if the latter are represented as graphs. 
The ubiquity of networks with skewed distributions of degree is not accidental. Such broad distributions indicate that there
is a whole hierarchy of node roles based on their degrees, going from a large majority of nodes with low degree to 
a small subset of nodes with high degree, or ``hubs''. The hubs have a fundamental role for the structure and dynamics 
of networks. Random SF networks with degree exponent $\lambda<3$ have so many hubs that a very small fraction of 
links (vanishing in the limit of infinite system size) is enough to keep a macroscopic fraction of nodes of the graph
in the same connected component, which can be equivalently stated by saying that the 
percolation threshold is zero~\cite{cohen00,newman01,pastor00,dorogovtsev08,vazquez04}.
In Ref.~\cite{radicchi09} we have already studied Achlioptas processes with PR on random SF networks. Here we present some more detailed
calculations and add significantly new material.

The networks are constructed as follows.
The starting point is a set of $N$ nodes and given 
degree sequence $\{k_1, k_2, \ldots , k_N \}$. The degrees of the sequence are taken from a 
power law distribution with exponent $\lambda$. We set the average degree $\langle k\rangle$ equal to $5$.
If links are placed randomly, the procedure can 
be carried out with the configuration model~\cite{molloy95}, i.e. by connecting randomly selected pairs of 
stubs adjacent to the nodes, until no more stubs are available. This is actually the procedure we have adopted for the RG model.
For PR, instead, at each iteration we pick two pairs of stubs and apply the PR to find 
which pair of stubs has to be eventually joined in a link (the PR applies as we have schematically illustrated in Fig.~\ref{fig1}). 

In the degree exponent's range $\lambda<3$ we will present only results referring to PR, due to the absence of 
a percolation threshold for RG. 
A remarkable result found independently in Refs.~\cite{cho09} and \cite{radicchi09} is that the percolation transition
of the Achlioptas process with PR has a non-vanishing threshold for $\lambda>\lambda_c\sim 2.3$ (Fig.~\ref{figthres}).

In Figs.~\ref{figSF2.5} and \ref{figSF2.8} we show the scalings at $p_c$ of $P$ and $S$ for
$\lambda=2.5$ and $2.8$, respectively. At variance with what we have seen in Sections~\ref{sec:2d} and \ref{sec:er}, here
the scaling of $P$ at $p_c$ is non-trivial, as $P$ decreases with $N$ as a power law in both cases. This appears inconsistent 
with the typical scenario of a discontinuous transition, which generally yields the trivial scaling we have observed in Figs.~\ref{fig:d2}c,
~\ref{fig3d}c and \ref{figER}c. We shall come back to this issue in Section~\ref{sec:concl}.
A clean power law scaling at $p_c$ is also found for $S$ (Figs.~\ref{figSF2.5}b and \ref{figSF2.8}b), although
we have seen that the same happens for explosive discontinuous transitions as well. 
\begin{figure}
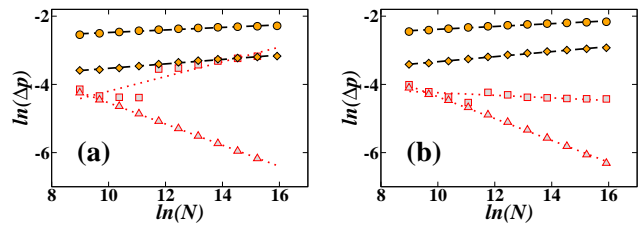

\includegraphics[width=0.22\textwidth,  height=0.16\textwidth]{ac_sf_g2.5_.eps}
\quad
\includegraphics[width=0.22\textwidth,  height=0.16\textwidth]{ac_sf_g2.8_.eps}
\caption{(Color online) Achlioptas test for SF networks. (a) $\lambda=2.5$: we plot the quantity $\Delta p$ 
as a function of the system size $N$. For both percolation models $\Delta p$ decreases as a power law, 
$\Delta p \sim N^{-\alpha}$, as $N$ increases. In particular we have: $\alpha=-0.04(1)$ (dashed black line) 
for RG (orange circles) and $\alpha=-0.26(3)$ (dotted red line) for PR 
(grey squares). We also consider the transition window $\Delta \tilde{p}$ (defined in Section~\ref{sec:analysis}), from which we obtain:  
$\tilde{\alpha}=-0.06(1)$ (lower black dashed line) for RG (orange diamonds); 
$\tilde{\alpha}=0.31(1)$ (lower red dotted line) for PR (grey triangles). (b) $\lambda=2.8$: 
same plot as the one of (a). The measured exponents are:  $\alpha=-0.04(1)$ (dashed black line) 
for RG (orange circles); $\alpha=0.04(1)$ (dotted red line) for PR (grey squares); 
$\tilde{\alpha}=-0.07(1)$ (lower black dashed line) for RG (orange diamonds); 
$\tilde{\alpha}=0.32(1)$ (lower red dotted line) for PR (grey triangles).}
\label{figAch2.5}
\end{figure}

In Fig.~\ref{fig:scaling_sf2.8} we show 
the rescaling of the variables $P$ and $S$. The data collapses observed in Figs.~\ref{fig:scaling_sf2.8}c and \ref{fig:scaling_sf2.8}f
show the profiles of the universal scaling functions $F^{(1)}$ and $F^{(2)}$ of Eqs.~(\ref{eqP}) and~(\ref{eqS}).
The results of the Achlioptas tests for $\lambda=2.5$ and $2.8$ are shown in Fig.~\ref{figAch2.5}. In each case 
we present the scaling of both $\Delta p$ and $\Delta\tilde{p}$, to check for the robustness of the results. 
We find that, while the scaling is clear for both variables, $\alpha\neq \tilde{\alpha}$. In fact, 
the exponents often indicate contradictory trends, with the transition window increasing and decreasing with $N$, which is clearly
inconsistent. We cannot exclude that this is due to finite size effects (as we have seen on lattices) 
and that simulations on much larger systems would 
show consistent results instead. On the other hand, it might be that 
the results of Achlioptas test indeed depend on the specific definition of the transition window.
\begin{figure}
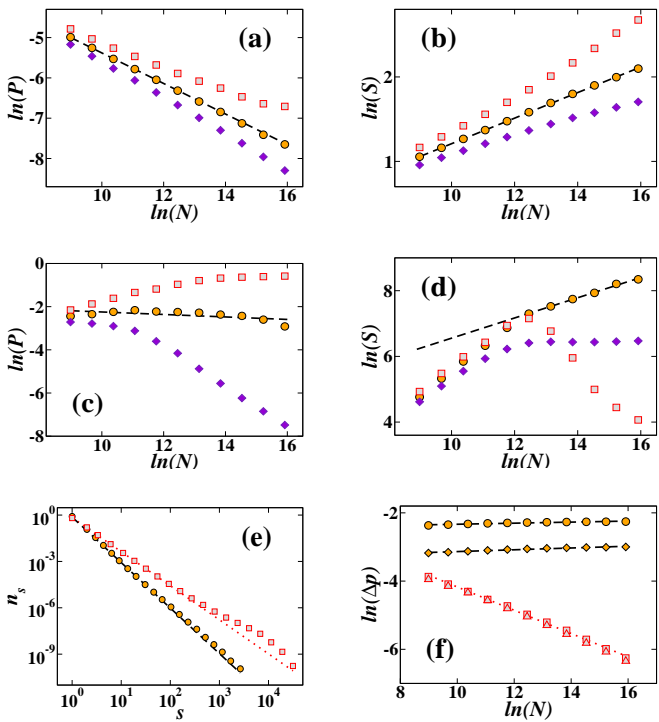

\includegraphics[width=0.22\textwidth,  height=0.16\textwidth]{beta_sf_g3.5_cl.eps}
\qquad
\includegraphics[width=0.22\textwidth,  height=0.16\textwidth]{gamma_sf_g3.5_cl.eps}

\vskip 0.4cm

\includegraphics[width=0.22\textwidth,  height=0.16\textwidth]{beta_sf_g3.5_pr.eps}
\qquad
\includegraphics[width=0.22\textwidth,  height=0.16\textwidth]{gamma_sf_g3.5_pr.eps}

\vskip 0.4cm

\includegraphics[width=0.22\textwidth,  height=0.16\textwidth]{tau_sf_g3.5.eps}
\qquad
\includegraphics[width=0.22\textwidth,  height=0.16\textwidth]{ac_sf_g3.5_.eps}

\caption{(Color online) Analysis of SF networks with degree exponent $\lambda=3.5$. (a) RG model: 
the percolation strength $P$ is plotted as a function of the system size $N$ for 
three different values of the occupation probability: $p=0.074$ (violet diamonds), 
$p=0.078$ (orange circles) and $p=0.082$ (grey squares). The dashed line stands for 
the best fit obtained at the critical point $p=p_c=0.078(1)$, which allows to determine 
$\beta/\nu=0.38(1)$. (b) RG model: the average cluster size $S$ is plotted as a function 
of the lattice side $L$ for the same values of $p$ used in (a). The dashed line has 
slope $\gamma/\nu=0.15(2)$. (c) PR model: the percolation strength $P$ is plotted 
as a function of the system size $N$ for three different values of the occupation 
probability: $p=0.2214$ (violet diamonds), $p=0.2224$ (orange circles) and $p=0.2234$ 
(grey squares). The dashed line stands for the best fit obtained at the critical point 
$p=p_c=0.2224(2)$, which yields $\beta/\nu=-0.06(3)$. (d) PR model:  
the average cluster size $S$ is plotted as a function of the network size $N$ for the 
same values of $p$ used in (c). The dashed line has slope $\gamma/\nu=0.40(9)$. (e)  
For both growth models $n_s \sim s^{-\tau}$ as $s$ increases. For RG (orange circles) 
$p_c=0.078(1)$ and  $\tau=2.94(1)$ (black dashed line), while for PR (grey squares) 
$p_c=0.2224(2)$ and $\tau=2.2(1)$ (red dotted line). Simulations have been performed 
on systems with $N=8192$. (f)  We plot the quantity $\Delta p$ as a function of the 
system size $N$. For both growth models $\Delta p$ decreases as a power law, 
$\Delta p \sim N^{-\alpha}$, as $N$ increases. In particular we have: 
$\alpha=-0.02(1)$ (upper dashed black line) for RG (orange circles) 
and $\alpha=0.34(1)$ (dotted red line) for PR (grey squares). 
We also consider the transition window $\Delta \tilde{p} = \tilde{p}_2-p_1$, where $\tilde{p}_2$ 
is the minimal value of the occupation probability at which $P=0.2$. In this 
case we find again a good power law fit. For RG (orange diamonds) the 
decay exponent is unchanged, since $\tilde{\alpha}=-0.02(1)$ (lower black dashed line). 
Similarly for PR (grey triangles) $\tilde{\alpha}=0.35(1)$.}
\label{fig3.5}
\end{figure}
\begin{table*}
\begin{tabular}{|c|c|c|c|c|c|c|c|}
\hline
System & Growth Model & $p_c$ & $\beta/\nu$ & $\gamma/\nu$ & $\tau$ & $\alpha$& $\tilde{\alpha}$ 
\\
\hline
\hline
\multirow{2}{*}{$2d$-lattice} & RG & $0.5$ & $0.11(1)$ & $1.76(1)$ & $2.05(1)$ & $0.15(1)$ & $0.16(1)$ 
\\
\cline{2-8}
 & PR & $0.5266(2)$ & $0.07(3)$ & $1.7(1)$ & $1.9(1)$ & $0.24(1)$ & $0.23(1)$ 
\\
\hline
\hline
\multirow{2}{*}{$3d$-lattice} & RG & $0.2488(3)$ & $0.48(1)$ & $2.0(1)$ & $2.20(1)$ & $0.10(1)$ & $0.10(1)$ 
\\
\cline{2-8}
 & PR & $0.3876(2)$ & $0.02(2)$ & $2.1(1)$ & $1.99(4)$ & $0.30(1)$ & $0.31(1)$ 
\\
\hline
\hline
\multirow{2}{*}{ER network} & RG & $0.5$ & $0.33(1)$ & $0.34(1)$ & $2.51(2)$ & $0.03(1)$ & $0.04(1)$ 
\\
\cline{2-8}
& PR & $0.8882(2)$ & $0.02(1)$ & $0.48(4)$ & $2.08(5)$ & $0.36(1)$ & $0.36(1)$
\\
\hline
\hline
\multirow{2}{*}{SF network $\lambda=2.5$} & RG & $0$ & $-$ & $-$ & $-$ & $-0.04(1)$ & $-0.06(1)$
\\
\cline{2-8}
& PR & $0.0629(1)$ & $0.59(1)$ & $0.24(1)$ & $2.15(2)$ & $-0.26(3)$ & $0.31(1)$ 
\\
\hline
\hline
\multirow{2}{*}{SF network $\lambda=2.8$} & RG & $0$ & $-$ & $-$ & $-$ & $-0.04(1)$ & $-0.07(1)$ 
\\
\cline{2-8}
& PR & $0.1329(1)$ & $0.50(1)$ & $0.42(1)$ & $2.13(6)$ & $0.04(1)$ & $0.32(1)$ 
\\
\hline
\hline
\multirow{2}{*}{SF network $\lambda=3.5$} & RG & $0.078(1)$ & $0.38(1)$ & $0.15(2)$ & $2.94(1)$ & $-0.02(1)$ & $-0.02(1)$ 
\\
\cline{2-8}
& PR & $0.2224(2)$ & $-0.06(3)$ & $0.40(9)$ & $2.2(1)$ & $0.34(1)$ & $0.35(1)$ 
\\
\hline
\end{tabular}
\caption{The table summarizes the results obtained from our numerical analysis. Percolation threshold and critical exponents 
are reported for each system and growth model analyzed.}
\label{table}
\end{table*}

For $\lambda>3$, however, the situation is different. Fig.~\ref{fig3.5} reports the results of our finite size 
scaling analysis for percolation transitions induced by RG and PR on SF networks with exponent $\lambda=3.5$.
In this case we also present the results of RG, because for $\lambda>3$ there is a non-zero threshold. 
In Ref.~\cite{cohen02} it has been proved that, for random percolation on SF networks, $\beta/\nu=1/(\lambda-1)$ and $\gamma/\nu=(\lambda-3)/(\lambda-1)$
for $3\leq\lambda\leq 4$. For $\lambda>4$, the process reaches the mean field limit and the exponents are frozen: $\beta/\nu=\gamma/\nu=1/3$. 
Interestingly, these are just the values of the exponents for the percolation transition of ER random networks. 
SF networks tend to ER random networks 
in the limit $\lambda\rightarrow\infty$. Our estimates of the exponents' ratios $\beta/\nu$ and $\gamma/\nu$ (Figs.~\ref{fig3.5}a and \ref{fig3.5}b)
are consistent with the predicted values for random percolation presented above.
For PR, instead, we recover the same scenario as on lattices and ER random networks. 
The scaling of $P$ at $p_c$ is trivial (Fig.~\ref{fig3.5}c), with $\beta/\nu=-0.06(3)$, which
is essentially zero, while the power law scaling of $S$ at $p_c$ is non-trivial (Fig.~\ref{fig3.5}d), with $\gamma/\nu=0.40(9)$. The Fisher exponent 
$\tau=2.2(1)$ (Fig.~\ref{fig3.5}e). The Achlioptas
test (Fig.~\ref{fig3.5}f) yields essentially the same set of values we had found for ER random networks (see Table~\ref{table}). Moreover, 
the values are stable no matter whether one uses $\Delta p$ or $\Delta \tilde{p}$.

\section{Discussion}
\label{sec:concl}

\begin{figure*}
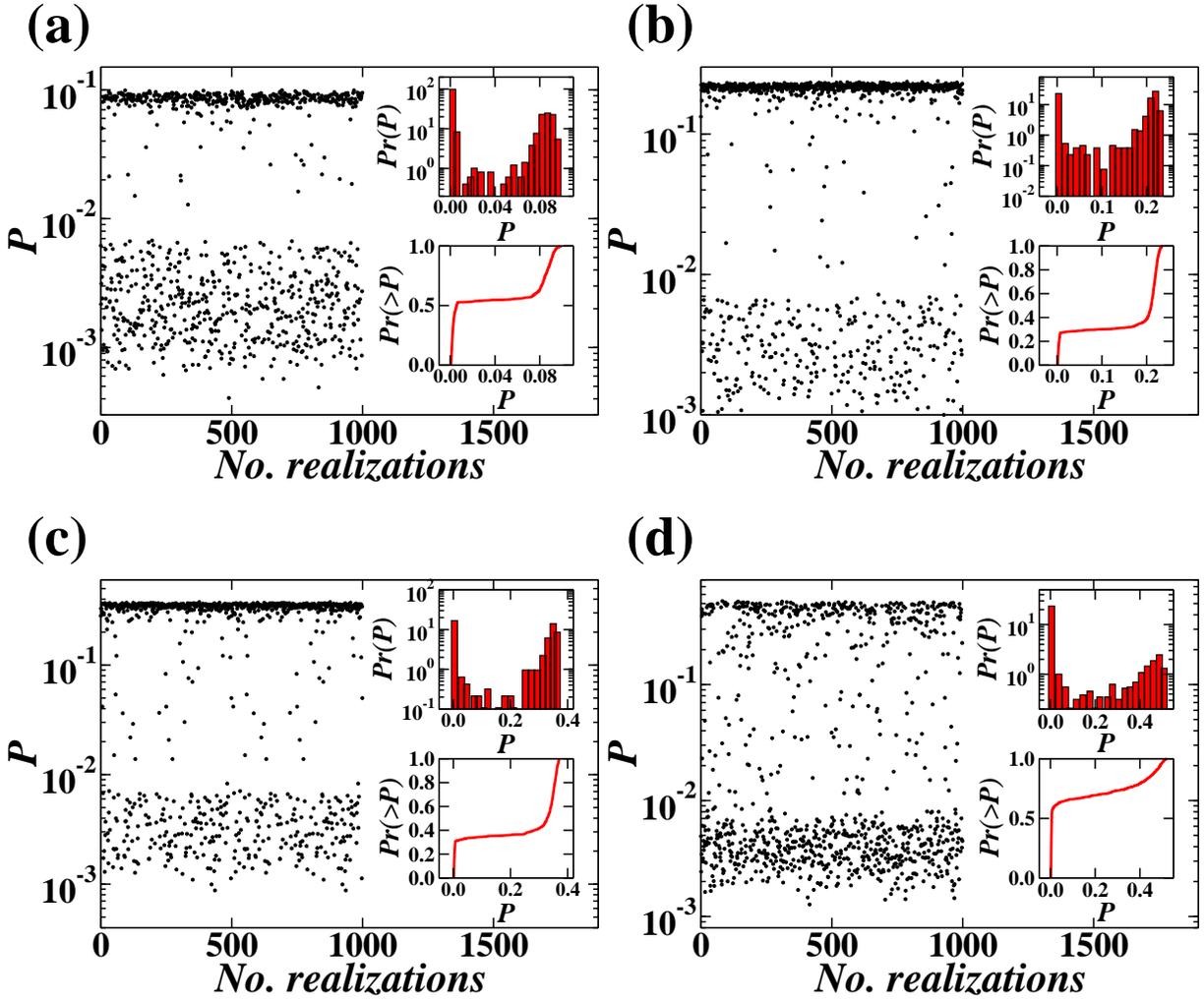

\begin{center}
\includegraphics[width=8cm]{pdistr_sf_g2.1_pr_pcN.eps}
\includegraphics[width=8cm]{pdistr_sf_g2.5_pr_pcN.eps}
\vskip0.4cm
\includegraphics[width=8cm]{pdistr_sf_g2.8_pr_pcN.eps}
\includegraphics[width=8cm]{pdistr_sf_g3.5_pr_pc.eps}
\end{center}
\caption{(Color online) Achlioptas process with PR on SF networks. Distributions of the values of the order parameter $P$ at the 
pseudocritical point $p_c(N)$ for different degree exponents $\lambda$: $2.1$ (a), $2.5$ (b), $2.8$ (c), $3.5$ (d). 
The main frame of each plot shows the values of $P$ for each of $1000$ realizations.
The insets display the distribution of the $P$-values (upper panel) and its cumulative (lower panel). The distributions are all bimodal,
which indicates that the order parameter undergoes a discontinuous jump at the critical point. The network size is $N=8192000$ in all cases.}
\label{fighist}
\end{figure*}
\begin{figure}
\includegraphics[width=\columnwidth]{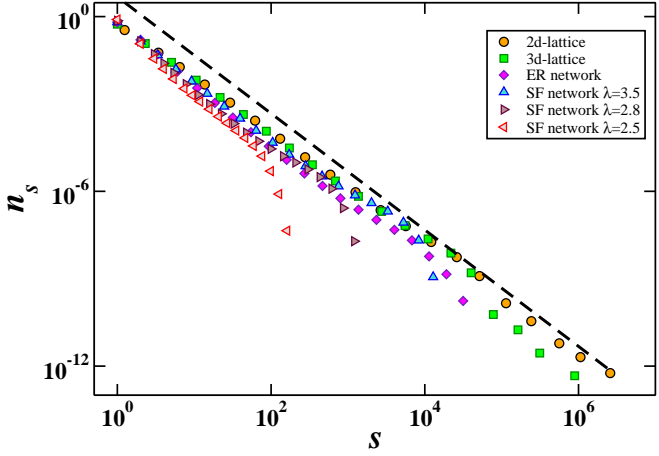}
\caption{(Color online) Cluster size distributions $n_s$ for Achlioptas processes with PR at the critical point. The cluster size distributions
scale as power laws (i.e., $n_s \sim s^{-\tau}$) for all systems analyzed in this paper. 
The Fisher exponents $\tau$ are
very close to each other and all distributions collapse into a unique curve with the only 
exceptions of the ones obtained for SF networks with $\lambda=2.5$ and $\lambda=2.8$. The dashed black line is a power law
with exponent $-2$, plotted as a useful reference.
The lattice side $L=4096$ for $2d$-lattice and $L=256$ for $3d$-lattice. $N=8192000$ for all networks.}
\label{tauall}
\end{figure}

In this section we want to discuss the results we have obtained, which are summarized in Table~\ref{table}. 
We have seen that our finite size scaling analysis leads to two different scenarios. The first scenario is consistent with 
the ``explosive'' transition
observed by Achlioptas {\it et al.}, and occurs on ER random networks, lattices and SF networks with 
degree exponent $\lambda>3$. In all these cases we have derived the same picture from finite size scaling, in particular
the saturation of the order parameter $P$ at $p_c$ with the size of the system $N$. On SF networks with $\lambda<3$
the situation looks different, as there we have observed a clear power law scaling of $P$ at $p_c$, just as one would expect
to find in continuous transitions. Moreover, the pseudo-critical points also show the clean power law scaling of Eq.~\ref{eq:chi2} for 
$\lambda<3$ (that is how the critical thresholds of Fig.~\ref{figthres} were derived), which usually happens for continuous transitions.
This appears to contradict the conclusion of Cho {\it et al.}, who claim that 
the transition is always discontinuous on SF networks~\cite{cho09}. Cho {\it et al.} have adopted the model by Chung and Lu~\cite{chung02}
to build their networks, which is different from the procedure we used, 
but we have verified that the results obtained in this way are consistent with ours. 

However, the seemingly continuous transition we observe for SF networks with $\lambda<3$ has the surprising and somewhat disturbing feature
that the hyperscaling relation of Eq.~\ref{hyper1} is violated, as one can easily verify through the values of $\beta/\nu$ and $\gamma/\nu$ 
reported in Table~\ref{table} for $\lambda=2.5$ and $\lambda=2.8$. Such violation could imply that the transition is not 
continuous after all. In order to test this, we have computed the distribution of the values of the order parameter $P$
at the pseudo-critical point $p_c(N)$, for PR on SF networks with $\lambda=2.1, 2.5, 2.8, 3.5$. The results are reported in Fig.~\ref{fighist}.
The two horizontal bands visible in the main frame of each of the four panels indicate that $P$ oscillates between two values at
the pseudo-critical point, which means that the transition is discontinuous. Interestingly, this is also found for $\lambda=2.1<\lambda_c$.
We could not carry out the finite size scaling analysis in this case, because the percolation threshold vanishes in the infinite size limit,
but the result on finite systems, as shown in Fig.~\ref{fighist}a, is the same as those for $\lambda>\lambda_c$. We conclude
that the percolation transition for an Achlioptas process with PR is discontinuous on SF networks, for any value of the degree exponent $\lambda$. 
Therefore, based on the results of this analysis, we have to partially modify the conclusion we had drawn in Ref.~\cite{radicchi09}, where we 
had stated that, for $\lambda<3$, the transition is continuous. There actually is a discontinuous jump of the order parameter at $p_c$:
nevertheless, all relevant percolation variables
display power law scaling at the percolation threshold for $\lambda<3$, in particular Eqs.~(\ref{eqP}),~(\ref{eqS}) and~(\ref{eq:chi2}) hold, just like in 
standard continuous transitions. Therefore, we hesitate to state that the transition is first- or second-order, as it looks like
an unusual mixture of both.
Therefore, the regime of SF networks for $\lambda<3$ is very intriguing and deserves further investigations.

Furthermore, the explosive transition observed in the other cases, including the original transition discovered by 
Achlioptas {\it et al.},
is not a standard discontinuous transition neither. The most striking feature here is that the size distribution 
$n_s$ of the ``finite'' clusters at $p_c$ is 
a power law, not exponential or Gaussian as one usually observes in first-order phase transitions. This fact has the consequence that all 
variables computed by means of $n_s$ also display non-trivial power law scaling at $p_c$, as we have seen with the average cluster size $S$.
Interestingly, the Fisher exponent $\tau$ for every transition we have investigated is very close to $2$, and consistent with 
this value within errors. In Fig.~\ref{tauall} we plot all distributions $n_s$ we have computed. Indeed, we see that the curves 
are strongly overlapping, and that only the curves corresponding to the anomalous 
discontinuous transition found in SF networks with $\lambda<3$ perhaps deviate from the general pattern, though very little.

\begin{figure}
\begin{center}
\includegraphics[width=\columnwidth]{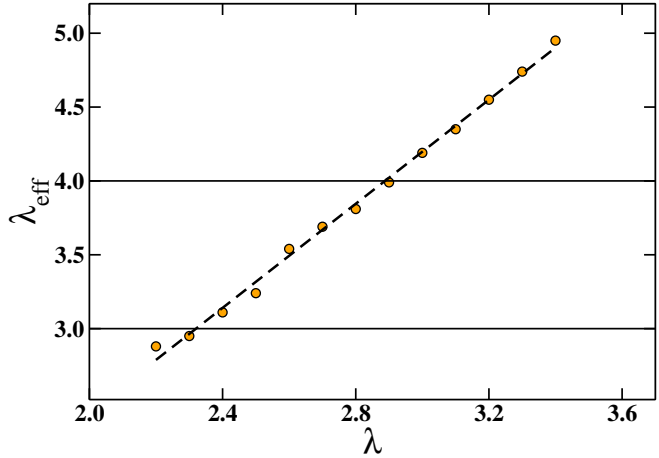}
\end{center}
\caption{(Color online) Achlioptas process with PR on SF networks.
Relation between the degree exponent $\lambda$ imposed through the starting degree sequence and the effective
exponent $\lambda_{eff}$ of the degree distribution of the system when it is at the percolation threshold. The 
relation is linear with good approximation (black dashed line). The newtwork size is $N=8192000$.}
\label{figexp}
\end{figure}

Another striking feature of our findings is the existence of a non-zero percolation threshold for SF networks 
for $\lambda>\lambda_c\sim 2.3$ (Fig.~\ref{figthres}), in contrast with the fact that the threshold for random percolation is zero until $\lambda=3$.
In Ref.~\cite{cho09}, Cho {\it et al.} suggested an interesting explanation
for this result. They noticed that, since in Achlioptas processes 
the networks are not constructed through the random addition of links, the degree distribution of the system during the growth 
deviates from that imposed by construction, which will be eventually reached at the end of the process. 
In Fig.~\ref{figexp} we plot the degree exponent $\lambda_{eff}$ measured at the critical point as a function of 
the imposed exponent $\lambda$. We see that the two exponents are quite different, and that there is a simple linear relation
between them. In particular, we notice that $\lambda_{eff}\sim 3$ when $\lambda=\lambda_c\sim 2.3$. Therefore, at the percolation threshold,
SF networks constructed with an Achlioptas process with PR for $\lambda>\lambda_c$ are actually SF networks with degree exponent 
bigger than $3$. For SF networks with degree exponent bigger than $3$, random percolation has a non-zero threshold, and this could be 
the reason of the non-zero threshold we observe for $\lambda>\lambda_c$. However, we stress that the ``effective'' SF networks
produced by Achlioptas processes are not random, so there is no {\it a priori} guarantee that one finds the same results
as for random SF networks as the argument above implies. Still, one could expect some qualitative agreement. 
This is confirmed by the fact that $\lambda_{eff}\sim 4$ when $\lambda=3$. For degree exponents larger than $4$, random SF networks
are hardly distinguishable from ER random networks. From the point of view of percolation the two classes of systems are in fact 
fully equivalent (same exponents). This could explain why, for $\lambda>3$, the picture we recover from finite size scaling looks
the same as for ER random networks, and that the corresponding critical exponents (including the exponent $\alpha$ of the Achioptas test) 
are consistent with each other within errors (see Table~\ref{table}).

\section{Summary}
\label{sec:summ}

In this paper we have performed a thorough numerical analysis of the percolation transitions induced
by Achlioptas processes with product rule. The typical outcome, on lattices, ER random networks and
SF networks is the ``explosive'' percolation transition originally observed by Achlioptas {\it et al.}~\cite{achlioptas09}.
This transition is kind of hybrid, as it combines the discontinuity of the order parameter at the critical point with analytical features
like the power law decay of the size distribution of finite clusters, a feature typical of continuous transitions. 
Hybrid phase transitions are actually not new, they have been observed in a variety of domains,
like spin glasses~\cite{gross84,kirkpatrick87}, 
constraint satisfaction problems (K-SAT)~\cite{monasson99} and models of jamming in granular materials~\cite{Ohern02,toninelli06,schwarz06}. 
A remarkable feature of our findings is that
the value of the exponent $\tau$ of the cluster size distribution appears to be compatible with $2$ in all instances, despite the diversity 
of the systems we considered.

For SF networks with degree exponent $\lambda<3$ the situation is even more extreme: on the one hand,
all percolation variables display power law scaling at the critical point, just like one expects for a continuous second-order phase transition;
on the other hand, the order parameter still undergoes a discontinuous jump at the critical point. 
This is certainly something worth investigation in the future. 
As usually in numerical studies of phase transitions, despite the 
large graph sizes we have investigated here, we cannot exclude that the regime we have tested is not yet ``asymptotic'' and that therefore
is dominated by finite size effects, which give a distorted perception of what truly happens. We tend to discard this 
hypothesis, though, due to the remarkably clean scaling plots we have derived.

Some theoretical arguments have been proposed to describe explosive percolation transitions~\cite{friedman09,moreira09}.
However, a real theory of such processes is still missing, and looking for a theory is certainly a challenging but promising future research 
direction. We hope that the results of our analysis will contribute to inspire new theoretical developments in this topic.

\begin{acknowledgments}
We would like to thank Antonio Coniglio, Sergey Dorogovtsev, Byungnam Kahng, Jinseop Kim, 
Jos\'e Fernando Mendes, Raissa D'Souza and Robert Ziff for stimulating discussions.
S. F. gratefully acknowledges ICTeCollective, grant number 238597 of the European Commission.
\end{acknowledgments}

\end{document}